\documentclass[english,11pt,a4paper]{article}
\usepackage{jheppub}
\usepackage[T1]{fontenc}
\usepackage[latin9]{inputenc}
\usepackage{float}
\usepackage{graphicx, hyperref}
\usepackage{amssymb,amsmath}
\usepackage{esint}
\usepackage{epstopdf}
\usepackage{babel}
\usepackage{amsxtra}

\numberwithin{equation}{section}

\usepackage{color}

\def\cB{{\mathcal B}}
\def\cR{{\mathcal R}}

\begin{document}

\title{Y-system for $Y=0$ brane in planar AdS/CFT}

\author[a,b]{Zolt\'an Bajnok, } 
\author[c]{Rafael I. Nepomechie, }  
\author[d]{L\'aszl\'o Palla } 
\author[e]{and Ryo Suzuki } 

 \affiliation[a]{
Theoretical Physics Research Group,
Hungarian Academy of Sciences, 1117 Budapest, P\'azm\'any s. 1/A
Hungary} 
\affiliation[b]{Institute for Advanced Studies, 
The Hebrew University of Jerusalem, 
Givat Ram Campus, 91904 Jerusalem, Israel
}
\affiliation[c]{
Physics Department, P.O. Box 248046,
University of Miami, Coral Gables, FL 33124, USA}
\affiliation[d]{
Institute for Theoretical Physics, Roland E\"otv\"os University,
1117 Budapest, P\'azm\'any s. 1/A Hungary}
\affiliation[e]{
Institute for Theoretical Physics and Spinoza Institute,
Utrecht University, 3508 TD Utrecht, The Netherlands}

{\hfill  UMTG-274, ITP-UU-12/15, SPIN-12/13}

{\hfill ITP-Budapest Report 660}

\emailAdd{bajnok@elte.hu}
\emailAdd{nepomechie@physics.miami.edu}
\emailAdd{palla@ludens.elte.hu}
\emailAdd{R.Suzuki@uu.nl}

\abstract{
The spectrum of open strings with integrable $Y=0$ brane boundary
conditions is analyzed in planar AdS/CFT. 
We give evidence that it can
be described by the same Y-system that governs the spectrum of closed
strings in ${\rm AdS}_5 \times {\rm S}^5$, except with different
asymptotic and analytical properties. We determine the asymptotic solution
of the $Y$-system that is consistent both with boundary asymptotic
Bethe ansatz and boundary L\"uscher corrections. 
}
\maketitle

\setcounter{footnote}{0}

\section{Introduction}

There have been recently immense interest and significant progress in
applying integrable methods to the planar AdS/CFT correspondence, see
\cite{Beisert:2010jr} and references therein.  The main focus has
concerned the spectral problem, which aims to determine the scaling
dimensions of gauge-invariant single-trace operators on one hand, and
the energy levels of closed strings on the other.  The sought-for
spectrum can be encoded into the $Y$-system of the problem, which,
when supplemented with the required asymptotical and analytical
information, provides the unique physical solution.

In this paper we focus on the extension of the spectral problem to a case 
with boundary.
Maximal giant gravitons \cite{McGreevy:2000cw, Grisaru:2000zn}
correspond to baryonic (or determinant) operators in ${\cal N}=4$ super Yang-Mills (SYM)
\cite{Balasubramanian:2001nh, Corley:2001zk}, and open strings ending
on a maximal giant graviton brane correspond to determinant-like
gauge-invariant operators \cite{Balasubramanian:2002sa,
deMelloKoch:2007uu}.  On the SYM side, Berenstein and V\'azquez
\cite{Berenstein:2005vf} found the Hamiltonian of an open spin chain
that gives the one-loop anomalous dimension of determinant-like
operators in the scalar sector.  They also determined the
corresponding boundary S-matrix (or reflection matrix), and showed
that it satisfies the boundary Yang-Baxter equation (BYBE)
\cite{Cherednik:1985vs, Sklyanin:1988yz, Ghoshal:1993tm}, suggesting
that the model is integrable.  (The double-row transfer matrix that
generates the Berenstein-V\'azquez Hamiltonian and the higher 
conserved charges was constructed and
diagonalized only recently in \cite{Nepomechie:2011nz}.) 
After some initial controversy \cite{Agarwal:2006gc, Okamura:2006zr}, Hofman
and Maldacena \cite{Hofman:2007xp} argued that integrability persists
at two loops.  As for bulk scattering, symmetry enhancement of the
asymptotic spin chain offers a guide to finding the all-loop boundary
S-matrix.  For the so-called $Y=0$ brane, which is the simplest and
most-studied example, $SU(1|2)^{2}$ symmetry determines the matrix
part of the boundary S-matrix \cite{Hofman:2007xp} (see
\cite{Ahn:2008df} for discussion on the BYBE).  The scalar factor was
found in \cite{Chen:2007ec} by solving the boundary crossing and
unitarity relations.  The corresponding all-loop asymptotic Bethe
ansatz (ABA) equations were studied in \cite{Galleas:2009ye}.  Based on
Yangian symmetry of boundary scattering \cite{Ahn:2010xa,
MacKay:2010ey}, bound-state boundary S-matrices were constructed in
\cite{Ahn:2010xa, Palla:2011eu}.  On the string theory side,
integrable boundary conditions for sigma models and their flat
connections were constructed in \cite{Mann:2006rh, Dekel:2011ja}.  For
a recent review including additional examples of boundaries, see
\cite{Zoubos:2010kh}.

The spectrum of open spin chains with finite length receives finite-size corrections,
and the predictions of the boundary ABA equations are no longer reliable.
The leading finite-size correction is due to virtual particles
reflecting between the two boundaries, and their contribution can be described 
by a L\"uscher-type formula \cite{Correa:2009mz, Bajnok:2010ui}.
For the exact description, the contributions from 
higher virtual processes have to be summed up. In the periodic case,
the sum of all virtual processes can be expressed by $Y$-functions 
which obey the $Y$-system.

The $Y$-system is a system of functional relations, which is related
to the symmetry of the problem \cite{Zamolodchikov:1991et, Kuniba:2010ir, Gromov:2010kf}.
It encodes the group-theoretical fusion hierarchy of the transfer
matrices in a gauge-invariant physical way.  Usually it can be derived
from an exact description of the problem, such as an integrable
lattice realization (see e.g. \cite{Bazhanov:1987zu, Kluemper1992304}) or exact
integral equations (TBA), which determine the finite-volume
ground-state energy.

In the AdS/CFT setting, the $Y$-system was conjectured
\cite{Gromov:2009tv} based on the experience in relativistic models
and by comparing its asymptotic solution to finite-size energy
corrections \cite{Bajnok:2008bm}.  Later it was derived for the ground
state from the thermodynamic Bethe ansatz equations
\cite{Bombardelli:2009ns, Gromov:2009bc, Arutyunov:2009ur,
Arutyunov:2009ux}.  Excited-state TBA equations obtained by analytical
continuation \cite{Dorey:1996re} lead to the same $Y$-system.
Although the scattering theory is invariant only under $SU(2\vert
2)^2$, the spectrum has the full $PSU(2,2\vert 4)$ symmetry.  Indeed,
it was shown in \cite{Gromov:2010vb} (see also
\cite{Arutyunov:2011uz}) that the conjectured $Y$-system exactly
corresponds to this symmetry.

Introducing integrable boundary conditions in a model usually
changes the asymptotic and analytical properties of the
$Y$-functions, but not the $Y$-system. This is true for integrable lattice 
models with boundaries (see e.g. \cite{Behrend:1995zj, OttoChui:2001xx}).
It was conjectured
and checked asymptotically that the $\beta$-deformed AdS/CFT
correspondence can be described by the undeformed $Y$-system
\cite{Gromov:2010dy}.  Later, using the model's realization in terms
of twisted boundary conditions \cite{Ahn:2010ws}, the $Y$-system and
the asymptotic and analytic information was derived from the
ground-state TBA equations \cite{Ahn:2011xq} 
(see also \cite{deLeeuw:2011rw}).

We expect that the $Y$-system used to describe the spectrum with
periodical and twisted boundary conditions will persist to the
boundary case with integrable boundary conditions.  This
expectation is also supported by the fact that the integrable boundary
condition corresponding to a Wilson loop leads via a boundary TBA
(BTBA) to the $Y$-system of $SU(2|2)$ \cite{Correa:2012hh,
Drukker:2012de}. In this paper we focus on a different integrable
boundary condition: the $Y=0$ brane \cite{Hofman:2007xp},
which describes the dimension of determinant-like SYM operators such as
\begin{equation}
{\cal O}_{Y}(Z^{k} \chi Z^{L-k})=\epsilon_{i_{1}\dots i_{N-1}i_{N}}^{j_{1}\dots j_{N-1}j_{N}}
Y_{j_{1}}^{i_{1}}\dots Y_{j_{N-1}}^{i_{N-1}}(Z^{k}\chi Z^{L-k})_{j_{N}}^{i_{N}} \,.
\end{equation}
Since in this case we do not have a BTBA equation for the ground state, we
simply assume that the $Y$-system is not changed, and check the
consistency of our assumption by direct computations of L\"uscher
corrections.  We can thereby determine the relevant asymptotic and
analytical solution, which is consistent with boundary L\"uscher and
asymptotic BA equations.

The paper is organized as follows: In the next Section
\ref{sec:Ysystem} we review the $Y$-system of the planar
$AdS_{5}/CFT_{4}$ correspondence.  Then in Section
\ref{sec:asymptotic} its asymptotic solution is determined in terms of
the eigenvalues of the double-row transfer matrices.  We construct a
generating functional from which the bound-state transfer matrix
eigenvalues can be extracted.  We use these quantities to calculate
the leading finite-size corrections of some operators and compare to
the literature with confirmation in Section \ref{sec:finitesize}.
Finally, we conclude in Section \ref{sec:conclusion}.  Some details of
the calculations, together with the implementation of the duality on
the asymptotic BA and transfer matrices, are relegated to the
Appendices.

\section{The $AdS_{5}/CFT_{4}$ Y-system}\label{sec:Ysystem}

The planar
$AdS_{5}/CFT_{4}$ correspondence can be described by an
integrable field theory, which has the global symmetry $PSU(2,2\vert4)$.
It is generally believed that the symmetry determines the Y-system \cite{Gromov:2009tv}, 
which, when supplemented with analyticity properties \cite{Cavaglia:2010nm,Balog:2011nm}, determines the
spectrum of the model. 

\begin{figure}[H]
\begin{centering}
\includegraphics[width=8cm]{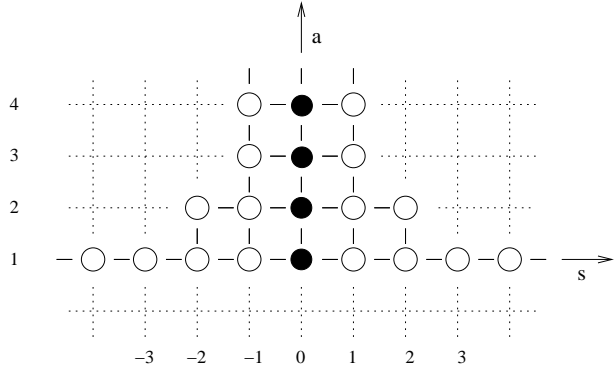}
\par\end{centering}

\caption{$Y$-system for planar 
AdS/CFT. The non-trivial $Y_{a,s}$-functions are marked
by circles, massive nodes with black. The vertical and horizontal
axes correspond to $a$ and $s$, respectively.}
\label{fig:Ysys}
\end{figure}

The form of the $Y$-system is very general 
\begin{equation}
\frac{Y_{a,s}^{+}Y_{a,s}^{-}}{Y_{a-1,s}Y_{a+1,s}}=\frac{(1+Y_{a,s+1})(1+Y_{a,s-1})}
{(1+Y_{a-1,s})(1+Y_{a+1,s})} \,,
\label{eq:Ysystem}
\end{equation}
and various models depend on the configurations of the nontrivial
$Y$-functions and the analytical properties in the generalized rapidity variable
$u$, $f^{\pm}(u)=f(u\pm\frac{i}{2})$.
The $PSU(2,2\vert4)$ symmetry 
of the planar
AdS/CFT integrable model leads to a T-shaped fat hook Y-system in Figure \ref{fig:Ysys}:
\footnote{In AdS/CFT setup, there are subtleties regarding the branch choice of $Y_{a,s}^\pm$ and 
the Y-system at $(a,s)=(2,\pm 2)$, which we neglect here.
The Y-system is equivalent to the TBA equations when these subtleties are correctly taken 
into account.
}

Models sharing the same symmetry often correspond to the same $Y$-system.
What is different is the analytical properties of the $Y$-functions.
The $AdS_{5}/CFT_{4}$ integrable model is more complicated than 
relativistic theories and has the unusual dispersion relation 
\cite{Beisert:2005tm,Dorey:2006dq}
\begin{equation}
E_{Q}(p_Q)=\sqrt{Q^{2}+16g^{2}\sin^{2}\frac{p_{Q}}{2}} \,,
\qquad g = \frac{\sqrt \lambda}{4 \pi} \,,
\end{equation}
where $Q\in\mathbb{Z}_{+}$ denotes the type of the particles: $Q=1$
corresponds to the fundamental particle, while $Q>1$ correspond 
to bound states of $Q$ fundamental particles. 
The rapidity parameter $u$ parameterizes the energy and momentum
as 
\begin{equation}
E_{Q}(u)=Q+2ig\left(\frac{1}{x^{[Q]}}-\frac{1}{x^{[-Q]}}\right) \,,
\qquad
p_{Q}(u)=-i\log\frac{x^{[Q]}}{x^{[-Q]}} \,,
\label{eq:Ep}
\end{equation}
where 
\begin{equation}
x(u)=\frac{u}{2g}+\sqrt{\frac{u}{2g}-1}\sqrt{\frac{u}{2g}+1} \,,\qquad
f^{[n]}(u)=f(u+\frac{in}{2}) \,.
\label{def:xu fpm}
\end{equation}
The momentum $p = p_Q (u)$ and the rapidity $u$ will be used interchangeably to parametrize physical quantities. 
The shifts $f^{\pm} = f^{[\pm 1]}$ are always understood to be with respect to the 
rapidity parameter.

The energy and momentum live on the torus parametrized by rapidity, while the $Y$-functions live on more complicated Riemann 
surfaces of rapidity.
The ground-state $Y$-functions can be constructed from the pseudo
energies of the mirror TBA equations \cite{Bombardelli:2009ns,Gromov:2009bc,Arutyunov:2009ur}.
The mirror model can be obtained from the original one by a double Wick rotation \cite{Ambjorn:2005wa,Arutyunov:2007tc},
$p\to -i\tilde{\epsilon}$, $E\to -i\tilde{p}$,
which amounts to using (\ref{eq:Ep}) with
\begin{equation}
x(u)=\frac{u}{2g}+i\sqrt{1-\frac{u^{2}}{4g^{2}}} \,.
\label{def:xu mir}
\end{equation}

With these kinematical variables, the energy of a fundamental multi-particle
state with momenta $p_{k}$ can be expressed in terms of only the
massive $Y$-functions, $Y_{Q}=Y_{Q,0}$, as 
\begin{equation}
E(L)=\sum_{k}E_{1}(p_{k})-\sum_{Q=1}^{\infty}\int\frac{du}{2\pi}\partial_{u}
\tilde{p}_{Q}\log(1+Y_{Q}) \,.
\end{equation}
The momenta are determined by the function $Y_{1}(p)$, analytically continued from \eqref{def:xu mir} to \eqref{def:xu fpm},
via the \emph{exact} Bethe equation: 
\begin{equation}
Y_{1}(p_{k})=-1 \,.
\end{equation}

We expect that this structure is valid for both the periodic and
the boundary situation. The difference lies in the asymptotic behavior
of the $Y$-functions, which we analyze in the next section. The integration
domains are also different. In the periodic case we integrate over
the whole line, while for the boundary case only over the half line.

\section{Asymptotic solution of the Y-system}\label{sec:asymptotic}

In this section we analyze the asymptotic large-volume solution of the $Y$-system. 

\subsection{The asymptotic solution in general}

The Y-system can be solved in terms of the $T$-system: 
\begin{equation}
T_{a,s}^{+}T_{a,s}^{-}=T_{a+1,s}T_{a-1,s}+T_{a,s+1}T_{a,s-1} \,,
\label{Tsystem}
\end{equation}
as 
\begin{equation}
Y_{a,s}=\frac{T_{a,s+1}T_{a,s-1}}{T_{a+1,s}T_{a-1,s}} \,.
\end{equation}
The $T$-functions are well-defined up to gauge transformations $T_{a,s}\to g^{[\pm a \pm s]}T_{a,s}$, 
where the signs are not correlated.
We shall look for the asymptotic
solution for large volume. In this limit the massive nodes are small
and the $T$-system of $PSU(2,2|4)$ splits into two copies of the $SU(2\vert2)$ $T$-systems
with boundary conditions $T_{a,0}=1$ \cite{Gromov:2009tv}. The small massive nodes at
leading order are determined by the asymptotic solutions of the two
$SU(2\vert2)$ wings as: 
\begin{equation}
Y_{a,0}=\frac{\phi^{[-a]}}{\phi^{[a]}}T_{a,-1}T_{a,1} \,.
\end{equation}
The unknown function $\phi$ can be fixed
by comparing it to the L\"uscher correction.
In the periodic case we obtain \cite{Bajnok:2008bm, Bajnok:2010ui}
\begin{equation}
Y_{a,0}=e^{-\tilde{\epsilon}_{a}L}\mathbb{T}_{a} \,,
\end{equation}
where $\mathbb{T}_{a}$ is an eigenvalue of the full transfer matrix with
the charge $a$ auxiliary representation space and the $N$-fold tensor
product of the fundamental representations, which by the usual abuse of notation we 
denote in the same way:
\begin{equation}
\mathbb{T}_{a}(p,\{p_{i}\})=\mbox{sTr}_{a}(\mathbb{S}_{aN}(p,p_{N})
\dots\mathbb{S}_{a1}(p,p_{1})) \,,
\end{equation}
where sTr means supertrace and $\mathbb{S}_{a j}$ denotes the full scattering matrix
of the charge $a$ auxiliary and the $j$-th fundamental particle.  
We introduce the basis for the fundamental representation of $SU(2\vert2)\otimes SU(2\vert2)$ by
\begin{equation}
| (\alpha \dot \alpha) \rangle = | \alpha \rangle \otimes |  \dot \alpha \rangle, \qquad
\alpha = 1,2,3,4, \qquad \dot \alpha = \dot 1, \dot 2, \dot 3, \dot 4.
\label{def:basis}
\end{equation}
Labels $1,2, \dot 1, \dot 2$ are bosonic, while $3,4,\dot 3, \dot 4$ are fermionic.

As the (fundamental) scattering matrix has a factorized $SU(2\vert2)\otimes SU(2\vert2)$
form
\begin{equation}
\mathbb{S}= S_{0}\, S\otimes\dot{S}, \qquad {\rm or} \qquad
\mathbb{S}_{(\alpha \dot \alpha)(\gamma \dot \gamma)}^{(\beta \dot \beta) (\delta \dot \delta)}
= S_{0}\, S_{\alpha \gamma}^{\beta \delta} \otimes\dot{S}_{\dot \alpha \dot \gamma}^{\dot \beta \dot \delta} \,,
\label{eq:S}
\end{equation}
the transfer matrix factorizes as well
\begin{equation}
\mathbb{T}_{a}=t_{a}\, t_{a,1}\otimes\dot{t}_{a,1}\,, \qquad
t_{a,1}(p\,;\{p_{i}\})=
\mbox{sTr}(S_{aN}(p,p_{N})\cdots S_{a1}(p,p_{1})) \,,
\end{equation}
and the normalization is 
\begin{equation}
t_{a}=t_{1}^{[1-a]}t_{1}^{[3-a]}\dots t_{1}^{[a-3]}t_{1}^{[a-1]}\,, 
\qquad
t_{1}=\prod_{i=1}^{N}S_{0}(p,p_{i}) \,.
\end{equation}
Comparing the asymptotic solution of the $T$-system to the L\"uscher
correction, we can conclude that the $SU(2\vert2)$ $T$-functions are
the left/right $SU(2\vert2)$ transfer matrices, and that 
$\frac{\phi^{-}}{\phi^{+}}=\left(\frac{x^{-}}{x^{+}}\right)^{L}t_{1}$.
Clearly, the fused $SU(2\vert2)$ transfer matrices $t_{a,1}$ satisfy
the $T$-system relation; and together with $\phi$,  provide the needed
asymptotic solution in the periodic case \cite{Gromov:2009tv}. 

\begin{figure}[t]
\begin{center}
\includegraphics[scale=0.6]{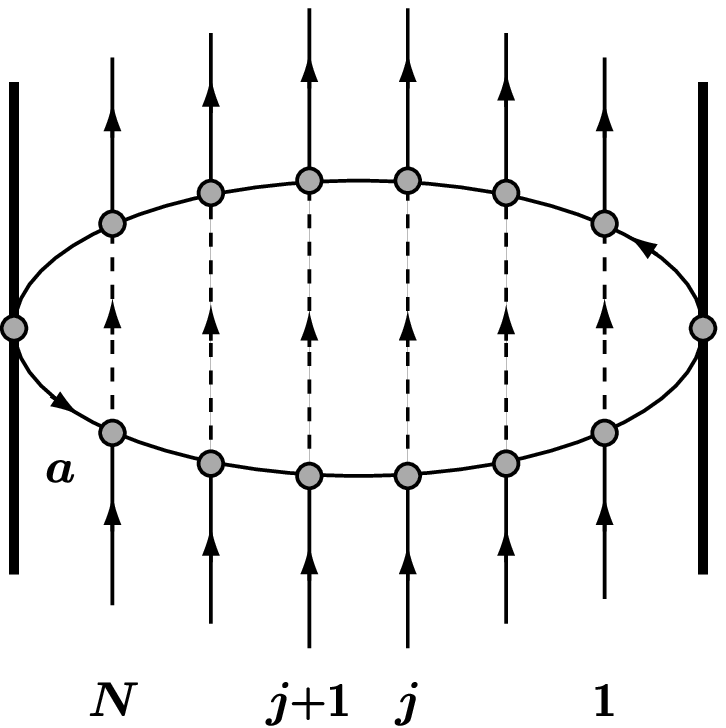}
\caption{The double-row transfer matrix \eqref{eq:Drow}. 
The inner indices, represented by dashed lines in the figure, must also be summed.}
\label{fig:drow}
\end{center}
\end{figure}

Let us now turn to the boundary case.  Comparing the asymptotic
solution with the boundary L\"uscher correction \cite{Bajnok:2010ui}, we
find 
\begin{equation}
Y_{a,0}=e^{-2\tilde{\epsilon}_{a}L}\mathbb{D}_{a} \,,
\end{equation}
where we have to replace the single-row transfer matrix with the double-row transfer
matrix \cite{Sklyanin:1988yz, Murgan:2008fs}:
\begin{equation}
\mathbb{D}_{a}(p,\{p_{i}\})=\mbox{Tr}_{a}(\mathbb{S}_{aN}(p,p_{N})\dots
\mathbb{S}_{a1}(p,p_{1})\mathbb{R}_{a}^{-}(p)\mathbb{S}_{1a}(p_{1},-p)
\dots\mathbb{S}_{Na}(p_{N},-p)\tilde{\mathbb{R}}_{a}^{+}(-p)) \,.
\label{eq:Drow}
\end{equation}
We remind the reader that $S_{aj}$ and $S_{ja}$ act nontrivially only on the 
vector spaces labeled by $a$ and $j$, and act as identity on all the other spaces; 
see Figure \ref{fig:drow}.
Here $\mathbb{R}_a^-(p)$ denotes the full reflection factor of the charge $a$ particle on the right 
boundary. Note that the transfer matrix is {\it not} written in 
terms of the left reflection factor
$\mathbb{R}_{a}^{+}(p) =  \mathbb{R}_{a}^{-}(-p)$, but instead
in terms of $\tilde{\mathbb{R}}_{a}^{+}(p)$. The latter is defined by
\begin{multline}
\mathbb{R}_{a}^{-}(p) \equiv \mbox{Tr}_{a'} {\cal P}_{a a'} 
\mathbb{S}_{a a'}(p,-p)\,  \tilde{\mathbb{R}}_{a'}^{+}(-p),
\\
{\rm or} \quad
\mathbb{R}_{a}^{-}(p)_{(\gamma \dot  \gamma)}^{(\beta \dot \beta)}=
\mathbb{S}_{aa}(p,-p)_{(\alpha \dot \alpha)(\gamma \dot \gamma)}^{(\beta \dot \beta)
 (\delta \dot \delta)} \tilde{\mathbb{R}}_{a}^{+}(-p)_{(\delta \dot \delta)}^{(\alpha \dot 
\alpha)}, 
\end{multline}
(where ${\cal P}$ is the permutation matrix), which
ensures that $\mathbb{D}_{a}(p_{j},\{p_{i}\})$ is equal to the 
boundary Bethe-Yang matrix \footnote{The boundary Bethe-Yang matrix 
is given (for the fundamental case) in (\ref{eq:BBYm}).}, and therefore 
$Y_{1,0}(p_j)=-1$ is equivalent to the boundary Bethe-Yang  
equations. (See appendix A in \cite{Ahn:2000jd} and \cite{Bajnok:2010ui} for further details).

Factorization of the reflection factors
\begin{equation}
\mathbb{R}^{-}=R_{0}^{-}R^{-}\otimes\dot{R}^{-}\,, \qquad 
\tilde{\mathbb{R}}^{+}=\tilde R_{0}^{+} \tilde{R}^{+}\otimes\dot{\tilde{R}}^{+}\,,
\label{eq:R}
\end{equation}
together with the factorization of the scattering matrix implies the
following factorization of the double-row transfer matrix
\begin{equation}
\mathbb{D}_{a}=d_{a}\, d_{a,1}\otimes\dot{d}_{a,1} \,,
\end{equation}
where
\begin{equation}
d_{a,1}(p\,;\{p_{i}\})= \mbox{Tr}_{a}(S_{aN}(p,p_{N})\cdots S_{a1}(p,p_{1})
R_{a}^{-}(p)
S_{1a}(p_{1},-p)\dots S_{Na}(p_{N},-p)\tilde{R}^+_{a}(-p)) \,,
\label{eq:da1}
\end{equation}
and the normalization is 
\begin{equation}
d_{a}=d_{1}^{[1-a]}d_{1}^{[3-a]}\dots d_{1}^{[a-3]}d_{1}^{[a-1]}\,, 
\qquad
d_{1}= R_{0}^{-}(p) \tilde R_{0}^{+}(-p) \prod_{i=1}^{N}S_{0}(p,p_{i})S_{0}(p_{i},-p) \,.
\end{equation}
Comparing the two expressions we can conclude that, in the boundary
setting, the asymptotic solution of the $T$-system is 
\begin{equation}
T_{a,1}=d_{a,1}\,, \qquad
T_{a,-1}=\dot{d}_{a,1}\,, \qquad
\frac{\phi^{-}}{\phi^{+}}=e^{-2\tilde{\epsilon}_{1}L}d_{1} \,.
\end{equation}
As the calculation of the bound-state transfer matrices starting from the 
definition is very cumbersome, we turn to their generating functional. The
generating functional is a compact solution of the $T$-system \cite{Kazakov:2007fy}
that is directly related to the fundamental transfer matrix. We start
by calculating the generating functional for the $su(2)$ sector in
the next subsection, and we then proceed with the general case.

\subsection{The fundamental double-row transfer matrix}

In this subsection we construct the fundamental $SU(2\vert2)$ double-row 
transfer matrix and explain its relation to the boundary asymptotic
Bethe ansatz equations. In so doing we first fix our conventions.
We normalize the fundamental scattering matrix (\ref{eq:S}) in the
$su(2)$ compatible way \cite{Arutyunov:2006yd, Arutyunov:2008zt}: 
\begin{equation}
S_{0}(x_{1},x_{2})=\frac{x_{1}^{+}+\frac{1}{x_{1}^{+}}-x_{2}^{-}-
\frac{1}{x_{2}^{-}}}{x_{1}^{-}+\frac{1}{x_{1}^{-}}-x_{2}^{+}-
\frac{1}{x_{2}^{+}}}\frac{x_{1}^{-}}{x_{1}^{+}}\frac{x_{2}^{+}}{x_{2}^{-}}
\sigma^{2}(p_{1},p_{2})\,, \qquad
S_{11}^{11}(x_{1},x_{2})=1=
S_{\dot{1}\dot{1}}^{\dot{1}\dot{1}}(x_{1},x_{2}) \,.
\end{equation}
The reflection factor on the right boundary (\ref{eq:R}) is simply
\begin{equation}
R^{-}(p)=\dot{R}^{-}(p)=\mbox{diag}(e^{-i\frac{p}{2}},-e^{i\frac{p}{2}},1,1) \,,
\qquad 
R_{0}^{-}(p)=-e^{-ip}\sigma(p,-p) \,.
\label{RightReflectionMatrix}
\end{equation}
The reflection factor on the left boundary is related to the right
one as
\begin{equation}
\mathbb{R}^{+}(p)=\mathbb{R}^{-}(-p) \,.
\label{LeftReflectionMatrix}
\end{equation}

Let us start to analyze a multiparticle state having particles of
type $1\dot{1}$ only; see \eqref{def:basis}.
The boundary asymptotic Bethe ansatz expresses
the single-valuedness of the wave function:
\begin{equation}
e^{-2ip_{j}(L+1)} \prod_{k=j-1}^{1}S_{0}(p_{j},p_{k})R_{0}^{-}(p_{j})
\prod_{k=1:k\neq j}^{N}S_{0}(p_{k},-p_{j})R_{0}^{-}(p_{j})
\prod_{k=N}^{j+1}S_{0}(p_{j},p_{k})=1 \,,
\label{eq:BBY1dot1}
\end{equation}
where the shift $L\rightarrow L+1$ is due to contributions from
the matrix parts of the reflection matrices. 
For more general states one has to diagonalize the boundary Bethe-Yang
matrix: 
\begin{equation}
\prod_{k=j-1}^{1}\mathbb{S}_{jk}(p_{j},p_{k})\mathbb{R}^{-}_{j}(p_{j})
\prod_{k=1:k\neq j}^{N}\mathbb{S}_{kj}(p_{k},-p_{j})\mathbb{R}^{+}_{j}(-p_{j})
\prod_{k=N}^{j+1}\mathbb{S}_{jk}(p_{j},p_{k}) \,.\label{eq:BBYm}
\end{equation}
Actually one has to diagonalize a family of such matrices obtained
by moving each particle {}``around'' the others by reflecting on
both boundaries. This is done at once by defining the double-row transfer
matrix of Sklyanin (\ref{eq:Drow}) with $a=1$, and there $\tilde{\mathbb{R}}_{1}^{+}(-p)$
was defined in such a way that $\mathbb{D}_{1}(p_{j},\{p_{i}\})$ gives
back the boundary Bethe-Yang matrix (\ref{eq:BBYm}). As both the
scattering and reflection matrices factorize, we focus on one copy
of the double-row transfer matrices. For concreteness,
we normalize them as 
\begin{equation}
\tilde{d}_{1,1}= \mbox{sTr}_{1}\left(S_{1N}(p,p_{N})\dots 
S_{11}(p,p_{1})R^{-}_{1}(p)S_{11}(p_{1},-p)
\dots S_{N1}(p_{N},-p)R^{-}_{1}(-p)\right),
\label{tilded11}
\end{equation}
which differs from $d_{1,1}$ (\ref{eq:da1}) since we used
$R^{-}(-p)$ instead of $\tilde{R}^{+}(-p)$. They are related to each other due
to the relation $R^{-}(-p)\propto (-1)^F \tilde{R}^{+}(-p)$, which changes the trace 
to supertrace. For later convenience, we record here that
\begin{equation}
\mathbb{D}_{1}(p) =\tilde{d}_{1}(p)\,\tilde{d}_{1,1}(p) 
\otimes\dot{\tilde{d}}_{1,1}(p) \,,
\label{eq:DD1}
\end{equation}
and note that the overall factor $\tilde{d}_{1}(p)$ will be determined in Section \ref{sec:finitesize}.

We first focus on the \emph{ground state} eigenvalue
of the transfer matrix $\tilde{d}_{1,1}(p)$ corresponding to 
$\vert1,1,\dots,1\rangle$.
We show in Appendix \ref{sec:vaceigenvalue} that the eigenvalue can be expressed
in terms of only the diagonal part as
\begin{equation}
\Lambda^{su(2)}(p)=\rho_{1}\Lambda_{1}+\rho_{2}\Lambda_{2}-\rho_{3}
\Lambda_{3}-\rho_{4}\Lambda_{4} \,,
\label{Lambda0}
\end{equation}
where 
\begin{eqnarray}
\Lambda_{1} & = & R_{\ \,\, 1}^{-1}(-p)S_{11}^{11}(p,p_{N})...
S_{11}^{11}(p,p_{1})R_{\ \,\, 1}^{-1}(p)S_{11}^{11}(p_{1},-p)... 
S_{11}^{11}(p_{N},-p)=1 \,,  \\
\Lambda_{2} & = & R_{\ \,\, 2}^{-2}(-p)S_{21}^{21}(p,p_{N})...
S_{21}^{21}(p,p_{1})R_{\ \,\, 2}^{-2}(p)S_{12}^{12}(p_{1},-p)... 
S_{12}^{12}(p_{N},-p)=\frac{\cR^{(-)+}}{\cR^{(+)+}}
\frac{\cB^{(-)-}}{\cB^{(+)-}}\,, \nonumber \\
\Lambda_{3} & = &\Lambda_{4} = R_{\ \,\, 3}^{-3}(-p)S_{31}^{31}(p,p_{N})
... S_{31}^{31}(p,p_{1})R_{\ \,\, 3}^{-3}(p)S_{13}^{13}(p_{1},-p)
... S_{13}^{13}(p_{N},-p)=\frac{\cR^{(-)+}}{\cR^{(+)+}} \nonumber \,,
\label{Lambdai}
\end{eqnarray}
and the functions $\cB^{(\pm)},\cR^{(\pm)}$ are defined in \eqref{def:BRQ mass}.
The rapidity-dependent $\rho$ functions are 
\begin{equation}
\rho_{1}=\frac{(1+(x^{-})^{2})(x^{-}+x^{+})}{2x^{+}(1+x^{+}x^{-})} \,,
\quad
\rho_{2}=\frac{x^{-}(x^{-}+x^{+})(1+(x^{+})^{2})}{2(x^{+})^{2}(1+x^{-}x^{+})} \,,
\quad
\rho_{3}+\rho_{4}=\frac{(x^{-}+x^{+})^{2}}{2(x^{+})^{2}} 
\,.
\label{rhos}
\end{equation}
As $\Lambda_{3}$ and $\Lambda_{4}$ are the same, only the combination
$\rho_{3}+\rho_{4}$ is determined.

Based on the analogy with the periodic theory \cite{Beisert:2005di, Beisert:2006qh}, we expect the generating
functional of the eigenvalues of the transfer matrices for anti-symmetric representations to be
of the form
\begin{eqnarray}
\tilde{\mathcal{W}}^{-1}&=&(1-\mathcal{D}\rho_{1}\Lambda_{1}\mathcal{D})(1-\mathcal{D}
\rho_{3}\Lambda_{3}\mathcal{D})^{-1}(1-\mathcal{D}\rho_{4}\Lambda_{4}
\mathcal{D})^{-1}(1-\mathcal{D}\rho_{2}\Lambda_{2}\mathcal{D})\nonumber \\ 
&=&
\sum_{a}(-1)^{a}\mathcal{D}^{a}\tilde{d}_{a,1}\mathcal{D}^{a}\,,
\label{eq:W}
\end{eqnarray}
where $\mathcal{D}=e^{-\frac{i}{2}\partial_{u}}$, and therefore $\mathcal{D} f = f^{-}\mathcal{D}$. 
In order to separate
$\rho_{3}$ and $\rho_{4}$, we demand that the state without particles
($\cB^{(\pm)}=\cR^{(\pm)}=1$) corresponds to the BPS state ${\cal O}_Y (Z^L)$, and thus
all higher transfer matrices, $\tilde{d}_{a,1}$, vanish.
This implies that 
\begin{equation}
\tilde{\mathcal{W}}=1\quad\longrightarrow\quad\rho_{1}=\rho_{3}\,, \qquad 
\rho_{2}=\rho_{4} \,.
\end{equation}
Let us renormalize the transfer matrices similarly to the periodic
case by dividing by $\rho_{3}\Lambda_{3}$ as: 
\begin{eqnarray}
\mathcal{W}_{su(2)}^{-1}&=&(1-\mathcal{D}\frac{\cR^{(+)+}}{\cR^{(-)+}}
\mathcal{D})(1-\mathcal{D}^{2})^{-1}(1-\mathcal{D}\frac{u^{+}}{u^{-}}
\mathcal{D})^{-1}(1-\mathcal{D}\frac{u^{+}}{u^{-}}\frac{\cB^{(-)-}}{\cB^{(+)-}}
\mathcal{D}) \nonumber \\
&=&\sum_{a}(-1)^{a}\mathcal{D}^{a}\hat{d}_{a,1}\mathcal{D}^{a} \,,
\label{eq:WW}
\end{eqnarray}
where we used that $\frac{\rho_{2}}{\rho_{1}}=\frac{\rho_{4}}{\rho_{3}}=\frac{u^{+}}{u^{-}}$
. The relation to $\tilde{d}_{a,1}$ is simply 
\begin{equation}
\tilde{d}_{a,1}=f^{[a-1]}f^{[a-3]}\dots f^{[3-a]}f^{[1-a]}
\hat{d}_{a,1}\,,
\qquad f=\rho_{3}\Lambda_{3} \,.
\label{tilde da1}
\end{equation}
Computing the generating functional, we found that 
\begin{equation}\label{gentrans}
(-1)^{a}\hat{d}_{a,1}=(a+1)\rho_{B1}-a\rho_{F1}\frac{\cR^{(+)[a]}}{\cR^{(-)[a]}}-
a\rho_{F2}\frac{\cB^{(-)[-a]}}{\cB^{(+)[-a]}}+(a-1)\rho_{B2}
\frac{\cR^{(+)[a]}}{\cR^{(-)[a]}}\frac{\cB^{(-)[-a]}}{\cB^{(+)[-a]}} \,,
\end{equation}
where 
\begin{equation}
\rho_{B1}=\rho_{B2}=\frac{u}{u^{[-a]}}\,, \qquad
\rho_{F1}=
\frac{u^{-}}{u^{[-a]}}\,, \qquad
\rho_{F2}=\frac{u^{+}}{u^{[-a]}} \,.
\end{equation}

As we did not derive (\ref{eq:W}), but merely conjectured, we performed
several consistency checks. First we analyzed $\hat{d}_{2,1}$. Using
the explicit form of the bound-state scattering matrices and reflection
factors we constructed the double-row transfer matrix $d_{a,1}$ of
(\ref{eq:da1}) for $a=2$ for $N=1, 2, 3$ particles at some 
randomly chosen momenta $p_{i}$ and coupling $g$. After 
verifying its commutativity properties,
we diagonalized it and compared its eigenvalue to $\hat{d}_{2,1}.$
After restoring the correct normalization factor we obtained perfect
agreement. We performed also another consistency check: we generated
the double-row transfer matrices for symmetric representations as 
\begin{equation}
\mathcal{W}_{su(2)}=\sum_{s}\mathcal{D}^s \hat{d}_{1,s}\mathcal{D}^s \,,
\end{equation}
and in a similar fashion we checked explicitly $\hat{d}_{1,2}$.

\subsection{Asymptotic Bethe ansatz and the generating functional}

We now turn to the analysis of generic states. Following 
\cite{Guan:1990xx, Galleas:2009ye}
and using experience with boundary systems, we expect the form of the
generic eigenvalue of the double-row transfer matrix to be of the
following dressed form: 
\begin{align}
\Lambda^{su(2)}=\left(\frac{x^{+}}{x^{-}}\right)^{m_{1}}\! \rho_{1}\frac{\cR^{(-)+}}{\cR^{(+)+}}
\Biggr[ \, &\frac{\cR^{(+)+}}{\cR^{(-)+}}\frac{\cB_{1}^{-}\cR_{3}^{-}}{\cB_{1}^{+}\cR_{3}^{+}}-
\frac{\cB_{1}^{-}\cR_{3}^{-}}{\cB_{1}^{+}\cR_{3}^{+}}\frac{Q_{2}^{++}}{Q_{2}}
\nonumber \\
&-\frac{u^+}{u^-}\frac{\cR_{1}^{+}\cB_{3}^{+}}{\cR_{1}^{-}\cB_{3}^{-}}\frac{Q_{2}^{--}}{Q_{2}}+
\frac{u^+}{u^-}\frac{\cB^{(-)-}}{\cB^{(+)-}}\frac{\cR_{1}^{+}\cB_{3}^{+}}{\cR_{1}^{-}\cB_{3}^{-}} \, \Biggl] \,,
\label{Lambda gen}
\end{align}
where the notation \eqref{def:BRQ mass}, \eqref{def:BRQ aux} is used.
Regularity of the transfer matrix at the roots gives the boundary
Bethe ansatz equations. Type $1$ roots are specified as $x^{+}(p)=y_i$,
type $2$ roots when $u=w_{l}$, and in the boundary case type $3$ roots
are equivalent to type $1$ roots: $x^{-}(p)=y_i^{-1}$. The corresponding Bethe equations read as
\begin{equation}
\frac{\cR^{(+)+}Q_{2}}{\cR^{(-)+}Q_{2}^{++}}\Bigg\vert_{x^{+}(p)=y_i}=1\,, 
\quad
\frac{u^-}{u^+}\frac{Q_{1}^{-} Q_{2}^{++}}{Q_{1}^{+} Q_{2}^{--}}
\Bigg\vert_{u=w_{l}}=-1\,, \quad
\frac{\cB^{(-)-}Q_{2}}{\cB^{(+)-}Q_{2}^{--}}
\Bigg\vert_{x^{-}(p)=y_i^{-1}}=1 \,.
\label{eq:BBA}
\end{equation}
Note that the equations for $y_i$ following from the first and third sets of equations in (\ref{eq:BBA}) are the same.
The second set of equations shows that the boundary factor $\frac{u^-}{u^+}$ can be removed formally by the redefinition $\tilde Q_1 (u) \equiv u \, Q_1 (u)$.

Let us describe their physical interpretation. Bethe ansatz equations diagonalize
the scatterings and reflections in terms of massive particles ($\bullet$)
and auxiliary {}``magnonic'' particles. In the $SU(2\vert2)$ problem
there are two types of magnons labelled by ${\sf y}$ and $\circ$. The
massive node scatters on the magnons as 
\begin{equation}
S_{\bullet {\sf y}}(p,y)=\frac{x^{-}-y}{x^{+}-y}\sqrt{\frac{x^{+}}{x^{-}}}\,, \quad
S_{{\sf y} \bullet}(y,p)=\frac{y-x^{+}}{y-x^{-}}\sqrt{\frac{x^{-}}{x^{+}}}\,,
\quad S_{\bullet\circ}(p,w)=1=S_{\circ\bullet}(w,p) \,,
\end{equation}
the remaining scattering matrices are
\begin{equation}
S_{\circ\circ}(w_{1},w_{2})=\frac{w_{1}-w_{2}-i}{w_{1}-w_{2}+i}\,, \quad
S_{{\sf y} \circ}(y,w)=S_{\circ {\sf y}}(y,w)=\frac{y+\frac{1}{y}-\frac{w}{g}+\frac{i}{2g}}{y+\frac{1}{y}-\frac{w}{g}-\frac{i}{2g}}\,, \quad 
S_{{\sf y} {\sf y}}(y_{1},y_{2})=1 \,.
\end{equation}
The boundary Bethe-Yang equations of the ${\sf y}$ magnons
with parameter $y_{k}$ are 
\begin{equation}
\prod_{j=1}^{N}S_{{\sf y} \bullet}(y_{k},p_{j})\prod_{l=1}^{m_{2}}S_{{\sf y} \circ}(y_{k},w_{l})
R_{\sf y}^{-}(y_{k})\prod_{l=1}^{m_{2}}S_{\circ {\sf y}}(w_{l},-y_{k})
\prod_{j=1}^{N}S_{\bullet {\sf y}}(p_{j},-y_{k})R_{\sf y}^{+}(-y_{k})=1 \,.
\end{equation}
Assuming that the reflection factors satisfy $R_{\sf y}^{-}(y)=R_{\sf y}^{+}(-y)$ and comparing to the Bethe
ansatz equations (\ref{eq:BBA}), we can conclude that these 
reflection factors are equal to 1
\begin{equation}
R_{\sf y}^{\pm}(y)=1 \,.
\end{equation}
In a similar fashion, we can write the equations for the magnons $\circ$
with rapidity $w_{k}$: 
\begin{equation}
\prod_{j=1:j\neq k}^{m_{1}}S_{\circ\circ}(w_{k},w_{j})\prod_{l=1}^{m_{2}}
S_{\circ {\sf y}}(w_{k},y_{l})R_{\circ}^{-}(w_{k})\prod_{l=1}^{m_{2}}S_{{\sf y} \circ}(y_{l},-w_{k})
\prod_{j=1:j\neq 
k}^{m_{1}}S_{\circ\circ}(w_{j},-w_{k})R_{\circ}^{+}(-w_{k})=1 \,.
\end{equation}
We assume also that the reflection factors satisfy $R_{\circ}^{-}(w)=R_{\circ}^{+}(-w)$ and compare
to the Bethe ansatz equation (\ref{eq:BBA}). Naively we would think
that $\frac{u^{-}}{u^{+}}$ corresponds to $R_{\circ}$; however,
the $k=j$ term in the product for $Q_{2}$ cancels it completely,
leading to
\begin{equation}
R_{\circ}^{\pm}(w)=1 \,.
\end{equation}
This is very similar to what has been obtained for the quark-antiquark
potential problem \cite{Drukker:2012de, Correa:2012hh}. 

The extension of the generating functional (\ref{eq:WW}) for generic
states reads as 
\begin{eqnarray}
\mathcal{W}_{su(2)}^{-1}&=&\left(1-\mathcal{D}\frac{\cR^{(+)+}}{\cR^{(-)+}}
\frac{\cB_{1}^{-}\cR_{3}^{-}}{\cB_{1}^{+}\cR_{3}^{+}}\mathcal{D}\right)\left(1-
\mathcal{D}\frac{\cB_{1}^{-}\cR_{3}^{-}}{\cB_{1}^{+}\cR_{3}^{+}}\frac{Q_{2}^{++}}{Q_{2}}
\mathcal{D}\right)^{-1}\qquad\qquad\qquad \label{eq:WWW}
\\ \nonumber &&
\qquad\qquad\times
\left(1-\mathcal{D}\frac{u^+}{u^-}\frac{\cR_{1}^{+}\cB_{3}^{+}}{\cR_{1}^{-}\cB_{3}^{-}}
\frac{Q_{2}^{--}}{Q_{2}}\mathcal{D}\right)^{-1}\left(1-\mathcal{D}\frac{u^+}{u^-}
\frac{\cB^{(-)-}}{\cB^{(+)-}}\frac{\cR_{1}^{+}\cB_{3}^{+}}{\cR_{1}^{-}\cB_{3}^{-}}\mathcal{D}\right) \\ \nonumber
&=& \sum_{a} (-1)^{a} \mathcal{D}^a \hat{d}_{a,1} \mathcal{D}^a \,.
\end{eqnarray}
The relation to $\tilde{d}_{a,1}$ is again given by \eqref{tilde da1}, 
except now
\begin{equation}
f= \rho_{1}\left(\frac{x^{+}}{x^{-}}\right)^{m_{1}}\frac{\cR^{(-)+}}{\cR^{(+)+}}\,.
\end{equation}
It is straightforward to check that the transfer matrices constructed from the generating 
functional satisfy the $T$-system of $SU(2|2)^2$ \eqref{Tsystem}.

\section{Checking the Y-functions: the boundary L\"uscher correction}\label{sec:finitesize}

We start by fixing the proper normalization of the double-row transfer
matrix $\mathbb{D}$. To this end, we analyze single particle states.
For a single particle the boundary Bethe-Yang equation reads as 
\begin{eqnarray}
1&=&e^{-2ipL}\mathbb{R}^{+}(-p)\mathbb{R}^{-}(p)=e^{-2ipL}\mathbb{R}^{-}(p)^{2} \\ \nonumber 
&=&e^{-2ip L} R_{0}^{-}(p)^{2}\mbox{diag}(e^{-ip},e^{ip},1,1)\otimes
\mbox{diag}(e^{-ip},e^{ip},1,1) \,,
\end{eqnarray}
where we have used (\ref{RightReflectionMatrix}) and (\ref{LeftReflectionMatrix}).
In particular, for the $(1\dot{1})$ particle, we obtain (see also 
(\ref{eq:BBY1dot1}))
\begin{equation}
1=e^{-2ip L} R_{0}^{-}(p)^{2} e^{-2ip} =
e^{-2ip(L+2)}\sigma^{2}(p,-p) \,,
\label{eq:1dot1}
\end{equation}
which at leading order leads to the momentum quantization
\begin{equation}
p_{1\dot{1}}=\frac{\pi}{L+2}n \,,
\label{eq:1dot12}
\end{equation}
where $n$ is an integer.
The analogous equations for the $(2\dot{2})$ particle are 
\begin{equation}
1=e^{-2ipL}\sigma^{2}(p,-p)\quad\longrightarrow\quad 
p_{2\dot{2}}=\frac{\pi}{L}n \,.
\label{eq:2dot21}
\end{equation}
Let us recover the same result from the double-row transfer matrix.
The 
fundamental transfer matrix with 
$N=1$ is given by (\ref{eq:DD1}), with
\begin{equation}
\tilde{d}_{1}(p) \equiv
S_0(p,p_1) S_0(p_1,-p)
\breve{R}_{0}^{+}(-p) R_{0}^{-}(p) \,,
\label{eq:tildeR0plus}
\end{equation}
where we have introduced the normalization factor 
$\breve{R}_{0}^{+}(-p)$ of the left reflection matrix that will be determined shortly.
The boundary Bethe-Yang equation from the double-row transfer matrix is 
\begin{equation}
\mathbb{D}_{1}(p_1,\{ p_1\}) e^{-2ip_{1}L}=-1 \,.
\end{equation}
Comparing with (\ref{eq:1dot1}), we see that 
\begin{equation}
\mathbb{D}_{1}(p, \{p\}) = - R_{0}^{-}(p)^{2} e^{-2ip} \,.  
\end{equation}    
Note that $\mathbb{D}_{1}$ is diagonal; and the only nonvanishing
contribution in the eigenvalue (\ref{Lambda0}) of $\tilde
d_{1,1}$ comes from the term with $\Lambda_{1}=1$.  Using also
that $S_{0}(p,p)=-1$ we obtain from (\ref{eq:tildeR0plus})
\begin{equation}
\breve{R}_{0}^{+}(-p)=\frac{e^{-2ip}R_{0}^{-}(p)}{S_{0}(p,-p)
\rho_{1}^{2}(p)} \equiv \frac{d_{0}(p)}{R_{0}^{-}(p)\rho_{1}^{2}(p)} \,,
\end{equation}
where in the second equality we have introduced the new quantity 
$d_{0}(p)$.
Having fixed the normalization of the left reflection factor, we now know
the properly normalized double-row transfer matrix.

\subsection{L\"uscher correction }

In the following we calculate the boundary L\"uscher correction of a
single impurity of type $(1\dot{1}$) and $(2\dot{2})$.  These two
states are in the $su(2)$ sector of the theory, and our transfer
matrices are devised to calculate the correction to their energy.
Correction to the energy of states of the form $(3\dot{3}$) or $(4\dot{4})$ can be easily
calculated from the eigenvalues of the dual transfer matrices,
which we obtain in Appendix C. The dualized transfer matrices are
also relevant for deriving the BTBA equations since the bound states
of the mirror theory are in the $sl(2)$ sectors.

The energies of the states $(1\dot{1}$) and $(2\dot{2})$ 
 are no longer degenerate because the residual symmetry of the $Y=0$ brane is $SU(1|2)^2$.
The properly normalized fundamental transfer matrix eigenvalue is 
(see (\ref{tilde da1}), (\ref{eq:tildeR0plus}))
\begin{equation}
\mathbb{D}_{1}=
f_{1,1}\hat{d}_{1,1}^{2} \,,
\end{equation}
where
\begin{equation}
f_{1,1} = \tilde{d}_{1} \left( \rho_3 \Lambda_3 \right)^2
= S_{0}(p,p_{1})S_{0}(p_{1},-p)d_{0}\left(
\frac{\cR^{(-)+}}{\cR^{(+)+}}\right)^{2} \,.
\end{equation}
Using the generating functional (\ref{eq:WW}) we can generate the antisymmetric
transfer matrices as 
\begin{equation}
\mathbb{D}_{a}=f_{a,1}\hat{d}_{a,1}^{2}\,, \qquad
f_{a,1}=
f_{1,1}^{[a-1]}f_{1,1}^{[a-3]}\dots f_{1,1}^{[3-a]}f_{1,1}^{[1-a]} \,.
\end{equation}
The L\"uscher correction in terms of this transfer matrix is 
\begin{equation}
\Delta E=-\sum_{a=1}^{\infty}\int\limits_0^\infty
\frac{dq}{2\pi} \, \mathbb{D}_{a} \, e^{-2 \tilde \epsilon_{a} L} 
\,,
\end{equation}
where $q$ is the mirror momentum of the $a$-th auxiliary particle.
This expression is exact at the leading order when the exponential
${\cal O} (e^{-2 \tilde \epsilon_{a} L} )$ is small, {\it e.g.} at
weak coupling.  We also expect that the L\"uscher $\mu$-term is absent
for fundamental particles at least in the weak-coupling limit
\cite{Bajnok:2008bm}.  We now calculate it at leading order and
compare to \cite{Bajnok:2010ui}.

\subsection{Weak-coupling expansion}

To make contact with the \lq\lq direct'' computation of \cite{Bajnok:2010ui} we use the parametrization
\footnote{Note that 
$z^{[+a]} +\frac{1}{z^{[+a]}} -\frac{ia}{2g}=\frac{q}{2g}+o(g)$ i.e. 
$u_{z_a}=\frac{q}{2}+o(g^{2})$.}
\begin{equation}
z^{[\pm a]}=\frac{q+ia}{4g}\left(\sqrt{1+\frac{16g^{2}}{q^{2}+a^{2}}}\pm1\right) \,.
\end{equation}
Then\footnote{Here $Q^{[a+1]}\equiv Q(\frac{q}{2})^{[a+1]}$ with $Q(u)$ defined in eq. \eqref{def:BRQ mass}.}
\begin{align*}
\frac{\cR^{(+)[a]}}{\cR^{(-)[a]}}&=\frac{z^{[a]}-x^{-}}{z^{[a]}-x^{+}}
\frac{z^{[a]}+x^{+}}{z^{[a]}+x^{-}}=\frac{(q-2u+{i}(a+1))}{(q-2u+
{i}(a-1))}\frac{(q+2u+{i}(a+1))}{(q+2u+{i}(a-1))}+
\dots=\frac{Q^{[a+1]}}{Q^{[a-1]}} \,,
\\
\frac{\cB^{(-)[-a]}}{\cB^{(+)[-a]}}&=\frac{1-z^{[-a]}x^{+}}{1-z^{[-a]}x^{-}}
\frac{1+z^{[-a]}x^{-}}{1+z^{[-a]}x^{+}}=\frac{(q-2u-{i}(a+1))}{(q-2u-
{i}(a-1))}\frac{(q+2u-{i}(a+1))}{(q+2u-{i}(a-1))}+
\dots=\frac{Q^{[-a-1]}}{Q^{[-a+1]}} \,.
\end{align*}
As $d_{0}=e^{-2ip}\frac{u^{-}}{u^{+}}=\left(\frac{z^{-}}{z^{+}}\right)^{2}\frac{u^{-}}{u^{+}}$
for $Q=1$, we find 
\begin{equation}
d_{0}^{[a-1]}d_{0}^{[a-3]}\dots d_{0}^{[3-a]}d_{0}^{[1-a]}=\left(
\frac{z^{[-a]}}{z^{[a]}}\right)^{2}\frac{u^{[-a]}}{u^{[a]}} \,.
\end{equation}
Similarly, 
\begin{equation}
\frac{\cR^{(-)+}}{\cR^{(+)+}}=\frac{(q-2u)(q+2u)}{(q-2u+i)(q+2u+i)}+
\dots\qquad\longrightarrow\qquad\frac{Q^{[1-a]}}{Q^{[1+a]}} \,,
\end{equation}
and for the scalar factor 
\begin{equation}
S_{0}(p,p_{1})S_{0}(p_{1},-p)=\left(\frac{z^{-}}{z^{+}} \right)^{2}
\frac{Q^{[2]}}{Q^{[-2]}}\quad\longrightarrow\quad\left( \frac{z^{[-a]}}{z^{[a]}} 
\right)^{2}\frac{Q^{[a-1]}Q^{[a+1]}}{Q^{[1-a]}Q^{[-1-a]}} \,.
\end{equation}
Collecting all factors, we obtain
\begin{equation}
f_{a,1} \simeq \left(  \frac{z^{[-a]}}{z^{[a]}} \right)^{4}\frac{u^{[-a]}}{u^{[a]}}
\frac{Q^{[a-1]}Q^{[1-a]}}{Q^{[a+1]}Q^{[-1-a]}}=\left(\frac{4g^{2}}{q^{2}+
{a^{2}}}\right)^{4}\frac{q-{i}a}{q+{i}a}
\frac{Q^{[a-1]}Q^{[1-a]}}{Q^{[a+1]}Q^{[-1-a]}} \,.
\end{equation}
From eq.(\ref{gentrans}) the contribution of the matrix part is 
\begin{equation}
\hat{d}_{a,1} \simeq \frac{(-1)^{a}2g^2}{q-ia}\left[\frac{2qa(q^{2}+a^{2}-1-4u^{2})(q^{2}
+a^{2}+1+4u^{2})}{(q^{2}+a^{2})(1+4u^{2})Q^{[a-1]}Q^{[1-a]}}\right] 
\,.
\end{equation}
Thus, the full contribution of the transfer matrix is 
\begin{equation}
\mathbb{D}_{a} \simeq \frac{(4g^{2})^6}{(q^{2}+a^{2})^5}\left[
\frac{qa(q^{2}+a^{2}-1-4u^{2})(q^{2}+a^{2}+1+4u^{2})}{(q^{2}+a^{2})
(1+4u^{2})}\right]^{2}\frac{1}{Q^{[a+1]}Q^{[-1-a]}Q^{[a-1]}Q^{[1-a]}} 
\,.
\end{equation}
To compute the L\"uscher correction we need the weak-coupling limit 
$e^{-2\tilde\epsilon_aL}=\left(\frac{4g^2}{q^2+a^2}\right)^{2L}$. Then, for
the $1\dot{1}$ particle, the leading L\"uscher correction is
\begin{equation}\label{egyegy}
\Delta E=-\sum_{a=1}^{\infty}\int\limits_0^\infty\frac{dq}{2\pi}
\frac{(4g^{2})^{2L+6}}{(q^{2}+a^{2})^{2L+5}}\left[
\frac{qa([q^{2}+a^{2}]^2-[1+4u^{2}]^2)}{(q^{2}+a^{2})
(1+4u^{2})}\right]^{2}\frac{1}{Q^{[a+1]}Q^{[-1-a]}Q^{[a-1]}Q^{[1-a]}} 
\,.
\end{equation}

For the $2\dot{2}$ particle the eigenvalue of the fundamental transfer matrix
is related very simply to that of the $1\dot{1}$ particle: repeating the
computation in  Appendix \ref{sec:vaceigenvalue}, it turns out that
\begin{equation}
\Lambda(p)_{2}^{su(2)}=e^{2ip}\left(
\rho_{1}\Lambda_{1}+\rho_{2}\Lambda_{2}-\rho_{3}
\Lambda_{3}-\rho_{4}\Lambda_{4}\right) \,.
\end{equation}
This means that the higher transfer matrices for the $2\dot{2}$ case differ
from the $1\dot{1}$ one only in their normalization
$\mathbb{D}_{a}\vert_{2\dot{2}}=\left(\frac{z^{[a]}}{z^{[-a]}}\right)^4\mathbb{D}_{a}\vert_{1\dot{1}}$. 
Therefore, in the weak-coupling limit, for the $2\dot{2}$ particle the 
leading L\"uscher correction can be written as
\begin{equation}\label{kettoketto}
\Delta E=-\sum_{a=1}^{\infty}\int\limits_0^\infty\frac{dq}{2\pi}
\frac{(4g^{2})^{2L+2}}{(q^{2}+a^{2})^{2L+1}}\left[
\frac{qa([q^{2}+a^{2}]^2-[1+4u^{2}]^2)}{(q^{2}+a^{2})
(1+4u^{2})}\right]^{2}\frac{1}{Q^{[a+1]}Q^{[-1-a]}Q^{[a-1]}Q^{[1-a]}} 
\,.
\end{equation} 

To evaluate these expressions for the L\"uscher corrections, we must set the $u$
parameter of the particle in question to the value that corresponds to one of
the momentum values allowed by the
boundary Bethe-Yang equations: $u=\cot(p_n/2)/2+o(g)$ as this guarantees that we
are dealing with an eigenvalue of the double-row transfer matrix. With these
$u$-s we compute the integrals in eq.(\ref{egyegy}, \ref{kettoketto}) by extending
the integration domain to the whole real line and using the residue theorem,
then we sum over $a$.

From \cite{Bajnok:2010ui} we know that $L=2$ is the smallest possible value 
among non-BPS operators $\sim {\cal O}_Y (Z \, \Phi^{a \dot a} \, Z^{L-1})$; in this
case for the $1\dot{1}$ particle we have (see \ref{eq:1dot12})
\begin{equation}
1\dot{1} \qquad L=2\,: \qquad 
p_n=\{\frac{\pi}{4},\frac{\pi}{2},\frac{3\pi}{4}\} \,.
\end{equation}
With these values we obtain
\begin{equation}
\label{egypiper2}
\Delta E_{1\dot{1}}(\frac{\pi}{2})=-2^6\cdot g^{20}\left(7\cdot
2^{5}\zeta(9)-429\cdot 2^2\zeta(13)+2431\zeta(17)\right) \,,
\end{equation}
and (see Appendix \ref{sec:trick})
\begin{eqnarray}
\Delta E_{1\dot{1}}(\frac{\pi}{4}) 
&=& -2^5\cdot g^{20}(-2^3\cdot 7\cdot
(99-70\sqrt{2})\zeta(9)-2(6765-4785\sqrt{2})\zeta(11) \nonumber \\
 && -2002(5\sqrt{2}-7)\zeta(15)+(7293-4862\sqrt{2})\zeta(17)) \,,\nonumber \\
\Delta E_{1\dot{1}}(\frac{3\pi}{4})
&=& -2^5\cdot g^{20}(-2^3\cdot 7\cdot
(99+70\sqrt{2})\zeta(9)-2(6765+4785\sqrt{2})\zeta(11) \nonumber \\
 && +2002(5\sqrt{2}+7)\zeta(15)+(7293+4862\sqrt{2})\zeta(17)) \,.
\label{egypiper4} 
\end{eqnarray}
  
For the $2\dot{2}$ particle with $L=2$ there is only one allowed $p$, 
namely $p=\frac{\pi}{2}$ (see \ref{eq:2dot21}), for which eq.(\ref{kettoketto}) gives 
\begin{equation}
\Delta E_{2\dot{2}}(\frac{\pi}{2})=-g^{12}\cdot 
2^6\left(21\zeta(9)-3\cdot 2^2\zeta(5)\right) \,.
\end{equation}
This expression coincides with the result of the direct L\"uscher calculation
based on the bound-state scattering and reflection matrices 
\cite{Bajnok:2010ui}\footnote{Note that the labeling of particles in this
paper and in \cite{Bajnok:2010ui} is different: what is called $2\dot{2}$
particle here is labeled as $1\dot{1}$ in \cite{Bajnok:2010ui} and vice
versa.
This can be seen by comparing the $SU(2\vert 2)$ reflections factors in the
two papers.}. 

For $L=3$ the boundary Bethe-Yang equations give
\begin{equation}
2\dot{2} \qquad L=3\,: \qquad 
p_n=\{\frac{\pi}{3},\frac{2\pi}{3}\} \,,
\end{equation}
and for the L\"uscher correction we find
\begin{equation}
\Delta E_{2\dot{2}}(\frac{\pi}{3})=
-g^{16}\cdot
2^3\left(15\zeta(7)-42\zeta(9)-165\zeta(11) +429\zeta(13)\right) \,.
\end{equation}
\begin{equation}
\Delta E_{2\dot{2}}(\frac{2\pi}{3})=
g^{16}\cdot
2^3\left(3645\zeta(7)+3402\zeta(9)-4455\zeta(11) -3861\zeta(13)\right) \,.
\end{equation}
Comparing to the analogous computation in the periodic case for the
Konishi operator \cite{Bajnok:2008bm}, it is interesting to observe  
that all of these L\"uscher corrections are given by linear combinations of
zeta functions, and there are no ``rational parts'' (i.e. terms
without zeta functions).
Although we have no explanation, this is perhaps
a generic feature of wrapping corrections for one-particle 
states (see also \cite{Fiamberti:2008sm,Fiamberti:2008sn,Gunnesson:2009nn,
Beccaria:2009hg,Gromov:2010dy}).
Also the leading L\"uscher correction of the $2\dot 2$ particle contains
smaller powers of $g$ than that of the $1\dot 1$, thus the \lq\lq wrapping
corrections'' to the anomalous dimension of the corresponding operators
appear in a lower loop order.

\section{Conclusion}\label{sec:conclusion}

We showed in this paper that the $Y$-system can be used to analyze the spectral 
AdS/CFT problem in the boundary setting. We identified the asymptotic solution which 
was consistent both with boundary asymptotic Bethe ansatz and boundary 
L\"uscher correction. Using this asymptotic solution we determined leading order 
wrapping corrections of various simple determinant-type operators. 

In our approach we assumed that the $Y$-system is independent of the
boundary condition, and indicated that its asymptotic solution
satisfies the possible requirements.  A more rigorous way would
be to derive the ground-state BTBA equations from first principles
({\it e.g.} string hypothesis), and prove that the $Y$-functions as constructed from the
pseudo-energies indeed satisfy the $Y$-system.  The solutions of the
$Y$-system equations have to satisfy excited-states BTBA equations.
It would be an interesting project to transform the $Y$-system into BTBA
equations based on the asymptotic solution that we have determined, 
similarly to the way it was done for the periodic case \cite{Cavaglia:2010nm,Balog:2011nm}.

We have explicitly constructed some bound-state double-row transfer matrices and checked that 
they indeed satisfy the functional relations of the $T$-system. We performed this task only for 
the first few transfer matrices where the bound-state scattering and reflection matrices 
were available. It would be nice to show decisively that they indeed satisfy the fusion hierarchy 
and can be analyzed in the fashion of \cite{Kazakov:2007fy}. 

In this paper we analyzed the $Y=0$ brane boundary condition. We expect that the same $Y$-system 
describes the finite-size spectrum of open strings with any other 
integrable boundary conditions. The asymptotic solutions could be extracted
from the asymptotic BA equations \cite{Correa:2009dm,Correa:2011nz}.

\section*{Acknowledgments}
We thank J\'anos Balog, Diego Correa and \'Arp\'ad Heged\H{u}s for 
useful discussions.
We thank the following institutions 
for hospitality during the course of this work: Centro de Ciencias de 
Benasque Pedro Pascual (ZB, RN, LP), Nordita (ZB, RN), University of 
Miami (ZB), and Roland E\"otv\"os University (RS).
This work is supported in part by Hungarian National Science Fund 
OTKA K81461 (ZB, LP); by the Institute for Advanced Studies,
Jerusalem, within the Research Group {\it Integrability and 
Gauge/String Theory} (ZB);
by the National Science Foundation under Grant PHY-0854366 
and a Cooper fellowship (RN); and by
the Netherlands Organization for Scientific Research (NWO) under the 
VICI grant 680-47-602 (RS).

\appendix

\section{Notation}

The $x$ variable is defined by
\begin{equation}
x(u)+\frac{1}{x(u)}=\frac{u}{g} \,,
\end{equation}
and it has branch points at $u=\pm 2g$.  As for the energy and
momentum of string states, we choose the branch cut along $u \in
(-2g,2g)$ as in \eqref{def:xu fpm}.  For the mirror particles, we
choose the branch cut along $u \in (-\infty,-2g) \cup (2g,
\infty)$ as in \eqref{def:xu mir}.  There are two conventions for the
mirror $x$ variable in the literature.  We adopt the choice ${\rm Im}
\, x > 0$ in this paper, as used in {\it e.g.} \cite{Bajnok:2008bm,
Gromov:2009tv}.  The other choice ${\rm Im} \, x < 0$ is used in {\it e.g.} 
\cite{Arutyunov:2009ur}.

The eigenvalues of double-row transfer matrix are expressed through
\begin{gather}
\cR^{(\pm)} =\prod_{i=1}^{N}\left(x(p)-x^{\mp}(p_{i})\right)\left(x(p)+
x^{\pm}(p_{i})\right)\,, \qquad
Q(u)=\prod_{i=1}^{N}(u-u_{i})(u+u_{i})\,,
\notag \\
\cB^{(\pm)} =\prod_{i=1}^{N}
\left(\frac{1}{x(p)}-x^{\mp}(p_{i})\right)\left(\frac{1}{x(p)}+
x^{\pm}(p_{i})\right)\,,
\label{def:BRQ mass}
\end{gather}
for the $N$-particle ground state $\vert1,1,\dots,1\rangle$, and
\begin{gather}
\cB_{1}\cR_{3}=\prod_{j=1}^{m_{1}}\left(x(p)-y_{j}\right)\left(x(p)+y_{j}\right)\,, \qquad
\cR_{1}\cB_{3}=\prod_{j=1}^{m_{1}}\left(\frac{1}{x(p)}-y_{j}\right)
\left(\frac{1}{x(p)}+y_{j}\right)\,,
\notag \\
Q_{1}(u) = \prod_{j=1}^{m_{1}} \left(\frac{u}{g}-y_j-\frac{1}{y_j}\right)\left(\frac{u}{g}+y_j+\frac{1}{y_j}\right)
= \left(\prod_{j=1}^{m_{1}} -\frac{1}{y_{j}^{2}} \right)
\cB_{1}\cR_{3}\cR_{1}\cB_{3} \,,
\notag \\
Q_{2}(u)= \prod_{l=1}^{m_{2}}(u-w_{l})(u+w_{l})\,,
\label{def:BRQ aux}
\end{gather}
for generic states with auxiliary roots: $2m_1$ is the number of $y$-roots, and $2m_2$ is the number of $w$-roots.

\section{Vacuum eigenvalue}\label{sec:vaceigenvalue}

The open-chain transfer matrix for a single copy of $SU(2|2)$ in the 
fundamental representation 
is given by (\ref{tilded11})  \cite{Sklyanin:1988yz, Murgan:2008fs}, 
which we now write as \footnote{Here 
the subscript $a$ denotes the 4-dimensional auxiliary space 
(fundamental representation).}
\begin{eqnarray}
\tilde{d}_{1,1}(p\,; \{p_{i}\}) = \mbox{sTr}_{a} {\cal M}_{a}(p\,; \{p_{i}\}) \,,
\label{transfer}
\end{eqnarray} 
where
\begin{eqnarray}
{\cal M}_{a}(p\,; \{p_{i}\})  = 
R_{a}^{+}(p)\, T_{a}(p\,; \{p_{i}\})\,
R_{a}^{-}(p)\,  \widehat T_{a}(p\,; \{p_{i}\}) \,,
\label{calM}
\end{eqnarray} 
and 
\begin{eqnarray}
T_{a}(p\,; \{p_{i}\}) &=& S_{a N}(p,p_{N}) \cdots S_{a 1}(p,p_{1}) \,,
\nonumber \\
\widehat T_{a}(p\,; \{p_{i}\}) &=& S_{1 a}(p_{1}, -p) \cdots S_{N 
a}(p_{N}, -p) .
\label{monodromy}
\end{eqnarray}
Here $S(p_{1},p_{2})$ is the non-graded $SU(2|2)$ bulk S-matrix
\cite{Beisert:2005tm} in the form given by \cite{Arutyunov:2008zt}.
We work in the ``string'' frame specified 
by (4.6) in \cite{Arutyunov:2008zt}. The right $Y=0$ boundary S-matrix 
$R^{-}(p)$ is given by the diagonal matrix 
(\ref{RightReflectionMatrix}) \cite{Hofman:2007xp, Ahn:2010xa}
\begin{eqnarray}
R^{-}(p) = \mathop{\rm diag}\nolimits(r^{-}_{1}\,,r^{-}_{2}\,, 1\,, 1 ) \,, \qquad
r^{-}_{1} = e^{-i p/2} \,,\quad r^{-}_{2} = -e^{i p/2} \,.
\end{eqnarray} 
The left boundary S-matrix 
$R^{+}(p)$ is given by
\begin{eqnarray}
R^{+}(p) = R^{-}(-p) = \mathop{\rm diag}\nolimits(r^{+}_{1}\,,r^{+}_{2}\,, 1\,, 1 ) \,, \qquad
r^{+}_{1} = e^{i p/2} \,,\quad r^{+}_{2} = -e^{-i p/2} \,.
\end{eqnarray} 
By construction, the transfer matrix has the fundamental 
commutativity property
\begin{eqnarray}
\left[ \tilde{d}_{1,1}(p\,; \{p_{i}\}) \,, \tilde{d}_{1,1}(p'\,; \{p_{i}\}) \right] = 0 
\label{commutativity}
\end{eqnarray}
for arbitrary values of $p$ and $p'$.

We choose the vector with all spins ``up'' as our vacuum state,
\begin{eqnarray}
|\Lambda^{(0)} \rangle =  \left( \begin{array}{c}
1 \\
0 \\
0 \\
0 \end{array} \right)^{\otimes N} \,.
\label{1vacuum}
\end{eqnarray} 
We shall now show that the corresponding vacuum eigenvalue is 
given by (\ref{Lambda0})

We begin by defining\footnote{In order to lighten the notation, we now refrain from 
writing the arguments.}
\begin{eqnarray}
{\cal U}_{a} = T_{a}\, R_{a}^{-}\,  \widehat T_{a} \,,
\end{eqnarray}
so that ${\cal M}_{a}$ defined in (\ref{calM}) is given by 
\begin{eqnarray}
{\cal M}_{a} = R_{a}^{+}\, {\cal U}_{a} \,.
\end{eqnarray} 
We see that the transfer matrix (\ref{transfer}) is given by
\begin{eqnarray} 
\tilde{d}_{1,1} &=& {\cal M}_{11} + {\cal M}_{22} - {\cal M}_{33} - {\cal M}_{44}\nonumber \\ 
&=& r^{+}_{1}\, {\cal U}_{11} + r^{+}_{2}\, {\cal U}_{22} - {\cal 
U}_{33} - {\cal U}_{44} \,,
\label{tcalU}
\end{eqnarray}
where the double-index subscripts of ${\cal M}$ and ${\cal U}$ denote matrix 
elements, regarding ${\cal M}$ and ${\cal U}$  as 
$4 \times 4$ matrices in the auxiliary space.

From the explicit form of $SU(2|2)$ S-matrix, we
now observe that $\widehat T_{a}$ acting on the vacuum gives
\begin{eqnarray}
\widehat T_{a} |\Lambda^{(0)} \rangle =
\left(\begin{array}{cccc}
\widehat T_{11} &\widehat T_{12} & \widehat T_{13} & \widehat T_{14} 
\\
0 & \widehat T_{22} & 0  & 0 \\
0 & \widehat T_{32} & \widehat T_{33}  & 0 \\
0 & \widehat T_{42} & 0  & \widehat T_{44}
\end{array}\right)  |\Lambda^{(0)} \rangle \,,
\label{hatTonvac}
\end{eqnarray}
and similarly for $T_{a}$ acting on the vacuum.
Hence,
\begin{eqnarray}
{\cal U}_{11} |\Lambda^{(0)} \rangle &=& r^{-}_{1}\, T_{11}\, \widehat 
T_{11} |\Lambda^{(0)} \rangle \nonumber \\
{\cal U}_{22} |\Lambda^{(0)} \rangle &=& \left( r^{-}_{1}\, T_{21}\, \widehat 
T_{12} + r^{-}_{2}\, T_{22}\, \widehat 
T_{22} +  T_{23}\, \widehat T_{32} +  T_{24}\, \widehat T_{4 2} 
\right)  |\Lambda^{(0)} \rangle \nonumber \\
{\cal U}_{33} |\Lambda^{(0)} \rangle &=& \left( r^{-}_{1}\, T_{31}\, \widehat 
T_{13} +  T_{33}\, \widehat T_{33} \right)  |\Lambda^{(0)} \rangle \nonumber \\
{\cal U}_{44} |\Lambda^{(0)} \rangle &=& \left( r^{-}_{1}\, T_{41}\, \widehat 
T_{14} +  T_{44}\, \widehat T_{44} \right) 
|\Lambda^{(0)} \rangle \,.
\label{calUonvac}
\end{eqnarray} 
The vacuum is an eigenstate of the diagonal elements of $T$ and 
$\widehat T$. We find that
\begin{eqnarray}
T_{ii}\, \widehat T_{ii} |\Lambda^{(0)} \rangle = 
 \widehat T_{ii}\, T_{ii} |\Lambda^{(0)} \rangle = \Lambda_{i} 
|\Lambda^{(0)} \rangle \,, \qquad i = 1, \ldots, 4 \,,
\end{eqnarray}
with corresponding eigenvalues $\Lambda_{i}$ given by (\ref{Lambdai}).

In order to deal with the terms in (\ref{calUonvac}) with off-diagonal
elements of $T$ and $\widehat T$, we exploit the commutation relations
that are encoded in the relation
\begin{eqnarray}
 T_{a}(p\,; \{p_{i}\})\,  S_{a b}(p,-p)\, \widehat T_{b}(p\,; \{p_{i}\}) 
=\widehat T_{b}(p\,; \{p_{i}\})\,  S_{a b}(p,-p)\, T_{a}(p\,; \{p_{i}\}) 
\,,
\label{hatTST}
\end{eqnarray}
which follows from the Yang-Baxter and unitarity equations. 
In particular, omitting terms that vanish when acting on the vacuum 
(see Eq. (\ref{hatTonvac})),
we have\footnote{These commutation relations 
correspond to the following matrix elements of (\ref{hatTST}) (viewed as a $16 
\times 16$ matrix equation in the auxiliary space):
(5,2), (7,10), (8,14), (3,9), (4,13), (9,3), (13,4), respectively.}
\begin{eqnarray}
\hspace{-0.4in}a_{1} T_{21}\, \widehat T_{12} + \frac{1}{2}(a_{1}-a_{2})  T_{22}\, \widehat T_{22}
+ a_{10} ( T_{23}\, \widehat T_{32} +  T_{24}\, \widehat T_{42})
&=& \frac{1}{2}(a_{1}-a_{2})  \widehat T_{11}\,  T_{11} \,, 
\label{(5,2)} \\
\hspace{-0.4in}a_{9} (T_{21}\, \widehat T_{12} +  T_{22}\, \widehat T_{22})
- a_{3}  T_{23}\, \widehat T_{32}  + \frac{1}{2}(a_{4}-a_{3})  T_{24}\, \widehat T_{42} 
&=& \frac{1}{2}(a_{1}-a_{2})  \widehat T_{31}\,  T_{13} +a_{9} 
\widehat T_{33}\,  T_{33}\,, \hspace{0.4in}\label{(7,10)} \\
a_{9} (T_{21}\, \widehat T_{12} +   T_{22}\, \widehat T_{22})
+ \frac{1}{2}(a_{4}-a_{3})   T_{23}\, \widehat T_{32}  - a_{3}  T_{24}\, \widehat T_{42} 
&=& \frac{1}{2}(a_{1}-a_{2})  \widehat T_{41}\,  T_{14} 
 +  a_{9} \widehat T_{44}\,  T_{44}
\,, \label{(8,14)} \\
a_{9} T_{11}\, \widehat T_{11}
&=& a_{1} \widehat T_{31}\,  T_{13} + a_{9} \widehat T_{33}\,  T_{33}\,, \label{(3,9)} \\
a_{9} T_{11}\, \widehat T_{11}
&=& a_{1} \widehat T_{41}\,  T_{14} + 
a_{9} \widehat T_{44}\,  T_{44}\,, \label{(4,13)} \\ 
a_{1} T_{31}\, \widehat T_{13} + a_{10} T_{33}\, \widehat T_{33}
&=& a_{10} \widehat T_{11}\,  T_{11}  \,, \label{(9,3)} \\
a_{1} T_{41}\, \widehat T_{14} + a_{10} T_{44}\, \widehat T_{44}
&=& a_{10} \widehat T_{11}\,  T_{11}  \,, \label{(13,4)}
\end{eqnarray} 
where $a_{i}= a_{i}(p,-p)$ are the matrix elements of $S_{a b}(p,-p)$ 
defined in \cite{Arutyunov:2008zt}.
We solve these equations for $T_{21}\, \widehat T_{12}$ in terms of 
the diagonal elements of $T$ and $\widehat T$, 
and obtain\footnote{We first combine (\ref{(5,2)}),  
(\ref{(7,10)}),  (\ref{(8,14)}) 
so as to cancel the terms $T_{23}\, \widehat T_{32} +  T_{24}\, 
\widehat T_{42}$; we then eliminate $\widehat  T_{31}\, 
T_{13}$ and $\widehat T_{41}\, 
T_{14}$ with (\ref{(3,9)}) and (\ref{(4,13)}), respectively; and we then
finally solve for $T_{21}\, \widehat T_{12}$.}
\begin{eqnarray}
T_{21}\, \widehat T_{12} |\Lambda^{(0)} \rangle = \left( c_{1} \Lambda_{1} + 
c_{2} \Lambda_{2} + c_{3} \Lambda_{3} \right) |\Lambda^{(0)} \rangle 
\,,
\end{eqnarray} 
where
\begin{eqnarray}
c_{1}&=&\frac{a_{1}-a_{2}}{2a_{1}} \,, \nonumber \\
c_{2}&=&\frac{a_{2}(3a_{3}-a_{4})+a_{1}(a_{4}-3a_{3})-8a_{9}a_{10}}{a_{1}(6a_{3}-2a_{4})+8a_{9}a_{10}}\,, \nonumber \\
c_{3}&=&\frac{2(a_{1}+a_{2})a_{9} a_{10}}{a_{1}\left[a_{1}(3 
a_{3}-a_{4})+ 4 a_{9}a_{10} \right]} \,.
\end{eqnarray}
It then follows from (\ref{(5,2)}) that
\begin{multline}
\lefteqn{( T_{23}\, \widehat T_{32} +  T_{24}\, \widehat T_{42}) 
|\Lambda^{(0)} \rangle}
\\
=\frac{1}{a_{10}}\Bigg\{
\left[\frac{1}{2}(a_{1}-a_{2})- a_{1}c_{1}\right]\Lambda_{1} +
\left[-\frac{1}{2}(a_{1}-a_{2})- a_{1}c_{2}\right]\Lambda_{2}
-a_{1}c_{3} \Lambda_{3} \Bigg\} |\Lambda^{(0)} \rangle \,.
\end{multline}

Denoting the vacuum eigenvalues of ${\cal U}_{ii}$ by $U_{ii}$, we now
see from (\ref{calUonvac}) that
\begin{eqnarray}
U_{11} &=& r^{-}_{1} \Lambda_{1} \,, \nonumber\\
U_{22} &=& r^{-}_{1} \left( c_{1} \Lambda_{1} + 
c_{2} \Lambda_{2} + c_{3} \Lambda_{3} \right)  + r^{-}_{2} 
\Lambda_{2} \nonumber \\
&&+ \frac{1}{a_{10}}\Bigg\{
\left[\frac{1}{2}(a_{1}-a_{2})- a_{1}c_{1}\right]\Lambda_{1} +
\left[-\frac{1}{2}(a_{1}-a_{2})- a_{1}c_{2}\right]\Lambda_{2} 
-a_{1}c_{3} \Lambda_{3} \Bigg\} \,, \nonumber \\
U_{33} &=& U_{44} = 
\frac{r^{-}_{1} a_{10}}{a_{1}}(\Lambda_{1}-\Lambda_{3}) + \Lambda_{3} 
\,.
\end{eqnarray} 
Finally, we see from (\ref{tcalU}) that the vacuum eigenvalue of the transfer 
matrix is given by
\begin{eqnarray}
\Lambda^{(0)} &=& r^{+}_{1}\, U_{11} + r^{+}_{2}\, U_{22} -2U_{33} 
\,, \nonumber \\
&=& \rho_{1} \Lambda_{1} + \rho_{2} \Lambda_{2} - (\rho_{3} + 
\rho_{4}) \Lambda_{3} \,,
\label{Lambda0raw}
\end{eqnarray} 
where
\begin{eqnarray}
\rho_{1} &=& r^{+}_{1} r^{-}_{1} +  r^{+}_{2}\left\{ 
r^{-}_{1} c_{1} + \frac{1}{a_{10}}\left[\frac{1}{2}(a_{1}-a_{2})- 
a_{1}c_{1}\right] \right\} - \frac{2 r^{-}_{1} a_{10}}{a_{1}}
\,, \nonumber \\
\rho_{2} &=& r^{+}_{2}\left\{  
r^{-}_{1} c_{2} + r^{-}_{2} 
+ \frac{1}{a_{10}}\left[-\frac{1}{2}(a_{1}-a_{2})- a_{1}c_{2}\right]
\right\}\,, \nonumber \\
\rho_{3} + \rho_{4} &=&  -r^{+}_{2} \left( r^{-}_{1} c_{3} - \frac{a_{1}c_{3}}{a_{10}} 
\right)
+2\left(1-\frac{r^{-}_{1} a_{10}}{a_{1}}\right) \,. 
\label{abcraw}
\end{eqnarray} 
By explicitly evaluating (\ref{abcraw}), we arrive at (\ref{rhos}),
and we see that (\ref{Lambda0raw}) coincides with (\ref{Lambda0}).
The eigenvalues corresponding to the other vacuum states can be 
computed in a similar way.

\section{Duality transformation}

In Appendix \ref{sec:vaceigenvalue} we computed the eigenvalue of the
double-row transfer matrix for the vacuum in the $su(2)$ sector.  The
$sl(2)$ sector is also often studied.  In this appendix we connect
the eigenvalues in these two sectors via a duality transformation on
the $y$-roots.

Following the standard procedure \cite{Beisert:2005di,Beisert:2005fw,Beisert:2006qh}, we dualize the $y$ roots in the double-row transfer matrix. 
The first equation of \eqref{eq:BBA} suggests the definition of
\begin{eqnarray}
q(x)=x^{2m_{2}}\left[\cR^{(+)}Q_{2}^{-}-\cR^{(-)}Q_{2}^{+}\right],
\label{q1}
\end{eqnarray}
which is a polynomial in $x$ of degree $2N+4m_{2}$. It has $m_{1}$
roots $y_{j}$ and $m_{1}$ roots $-y_{j}$. Hence, it additionally
has $\tilde{m}_{1}$ roots $\tilde{y}_{j}$ and $\tilde{m}_{1}$ roots
$-\tilde{y}_{j}$, where 
\begin{eqnarray}
\tilde{m}_{1}=N+2m_{2}-m_{1}\,.
\end{eqnarray}
 Factoring this polynomial, we obtain 
\begin{eqnarray}
q(x)=\gamma \cB_{1}\cR_{3}\tilde{\cB}_{1}\tilde{\cR}_{3}\,,\label{q2}
\end{eqnarray}
 where 
\begin{eqnarray}
\tilde{\cB}_{1}\tilde{\cR}_{3} = \prod_{j=1}^{\tilde{m}_{1}}\left(x(p)-
\tilde{y}_{j}\right)\left(x(p)+\tilde{y}_{j}\right),
\quad
\tilde{\cR}_{1}\tilde{\cB}_{3} = \prod_{j=1}^{\tilde{m}_{1}}\left(\frac{1}{x(p)}-
\tilde{y}_{j}\right)\left(\frac{1}{x(p)}+\tilde{y}_{j}\right),
\end{eqnarray}
and $\gamma$ is some nonzero constant. Forming the ratio of the
expressions (\ref{q1}) and (\ref{q2}), we see that 
\begin{eqnarray}
F(x)\equiv\frac{x^{2m_{2}}}{\cB_{1}\cR_{3}\tilde{\cB}_{1}\tilde{\cR}_{3}}
\left[\cR^{(+)}Q_{2}^{-}-\cR^{(-)}Q_{2}^{+}\right]
\end{eqnarray}
 is in fact independent of $x$. The identity $F^{+}=F^{-}$ implies
that \begin{eqnarray}
\frac{\cB_{1}^{-}\cR_{3}^{-}}{\cB_{1}^{+}\cR_{3}^{+}}\left[1-\frac{\cR^{(-)+}}{\cR^{(+)+}}
\frac{Q_{2}^{++}}{Q_{2}}\right]=\left(\frac{x^{-}}{x^{+}}\right)^{2m_{2}}
\frac{\tilde{\cB}_{1}^{+}\tilde{\cR}_{3}^{+}}{\tilde{\cB}_{1}^{-}\tilde{\cR}_{3}^{-}}
\left[\frac{\cR^{(+)-}}{\cR^{(+)+}}\frac{Q_{2}^{--}}{Q_{2}}-\frac{\cR^{(-)-}}{\cR^{(+)+}}
\right]\,.\label{id1}
\end{eqnarray}
Similarly, the identity $F(\frac{1}{x^{+}(p)})=F(\frac{1}{x^{-}(p)})$
implies that 
\begin{eqnarray}
\frac{\cR_{1}^{+}\cB_{3}^{+}}{\cR_{1}^{-}\cB_{3}^{-}}\left[\frac{Q_{2}^{--}}{Q_{2}}-
\frac{\cB^{(-)-}}{\cB^{(+)-}}\right]=\left(\frac{x^{-}}{x^{+}}\right)^{2m_{2}}
\frac{\tilde{\cR}_{1}^{-}\tilde{\cB}_{3}^{-}}{\tilde{\cR}_{1}^{+}\tilde{\cB}_{3}^{+}}
\left[\frac{\cB^{(+)+}}{\cB^{(+)-}}-\frac{\cB^{(-)+}}{\cB^{(+)-}}\frac{Q_{2}^{++}}{Q_{2}}
\right]\,.\label{id2}
\end{eqnarray}
With the help of the identities (\ref{id1}) and (\ref{id2}), the
expression \eqref{Lambda gen} for the eigenvalue becomes
\begin{align}
\Lambda^{sl(2)} = \left(\frac{x^{+}}{x^{-}}\right)^{m_{1}-2m_{2}}\frac{\cR^{(+)-}}{\cR^{(+)+}}
\rho_1\Bigg[ \, &\frac{\tilde{\cB}_{1}^{+}
\tilde{\cR}_{3}^{+}}{\tilde{\cB}_{1}^{-}\tilde{\cR}_{3}^{-}}
\frac{Q_{2}^{--}}{Q_{2}}- \frac{\cR^{(-)-}}{\cR^{(+)-}}
\frac{\tilde{\cB}_{1}^{+}\tilde{\cR}_{3}^{+}}{\tilde{\cB}_{1}^{-}
\tilde{\cR}_{3}^{-}}
\notag \\
&-\frac{u^+}{u^-} \frac{\cB^{(+)+}}{\cB^{(-)+}}
\frac{\tilde{\cR}_{1}^{-}\tilde{\cB}_{3}^{-}}{\tilde{\cR}_{1}^{+}
\tilde{\cB}_{3}^{+}}+ \frac{u^+}{u^-}\frac{\tilde{\cR}_{1}^{-}
\tilde{\cB}_{3}^{-}}{\tilde{\cR}_{1}^{+}\tilde{\cB}_{3}^{+}}\frac{Q_{2}^{++}}{Q_{2}} \, \Bigg]\,,\label{duallambda}
\end{align}
which corresponds to the $sl(2)$ grading. To obtain this form  
we used the identities 
\begin{eqnarray}
\frac{\cR^{(-)+}}{\cR^{(+)+}}\frac{\cB^{(-)+}}{\cB^{(+)-}} &=&
\frac{\cR^{(+)-}}{\cR^{(+)+}}\left[\frac{\cR^{(-)+}}{\cR^{(+)-}}
\frac{\cB^{(-)+}}{\cB^{(+)-}}\right]=\frac{\cR^{(+)-}}{\cR^{(+)+}}\,,\nonumber \\
\frac{\cR^{(-)+}}{\cR^{(+)-}}\frac{\cB^{(+)+}}{\cB^{(+)-}} &=& \left[\frac{\cR^{(-)+}}{\cR^{(+)-}}
\frac{\cB^{(-)+}}{\cB^{(+)-}}\right]\frac{\cB^{(+)+}}{\cB^{(-)+}}=\frac{\cB^{(+)+}}{\cB^{(-)+}}.
\end{eqnarray}

Apart from its normalization 
the expression in (\ref{duallambda}) is very similar to what is 
obtained in \cite{Gromov:2010dy, Ahn:2010ws}
for the dualized eigenvalue of the fundamental transfer matrix in the
{\sl twisted} periodic case; the $\frac{u^+}{u^-}$ functions play the role of
the -- now momentum dependent -- twist factors and the
$\tilde{\cB}_{1}\tilde{\cR}_{3}$ ($\tilde{\cR}_{1}\tilde{\cB}_{3}$) polynomials are
twice as long as in the periodic case.   
Based on this analogy from the expression in \cite{Gromov:2010dy} we expect  
the dual generating functional to be
\begin{multline}
\mathcal{W}_{sl(2)}^{-1}=\left(1-\mathcal{D}\frac{\cR^{(-)-}}{\cR^{(+)-}}
\frac{\tilde{\cB}_{1}^{+}\tilde{\cR}_{3}^{+}}{\tilde{\cB}_{1}^{-}\tilde{\cR}_{3}^{-}}\mathcal{D}\right)^{-1}\left(1-
\mathcal{D}\frac{\tilde{\cB}_{1}^{+}\tilde{\cR}_{3}^{+}}{\tilde{\cB}_{1}^{-}\tilde{\cR}_{3}^{-}}\frac{Q_{2}^{--}}{Q_{2}}
\mathcal{D}\right) \\
\qquad\qquad\times\left(1-\mathcal{D}\frac{u^+}{u^-}\frac{\tilde{\cR}_{1}^{-}\tilde{\cB}_{3}^{-}}{\tilde{\cR}_{1}^{+}\tilde{\cB}_{3}^{+}}
\frac{Q_{2}^{++}}{Q_{2}}\mathcal{D}\right)\left(1-\mathcal{D}\frac{u^+}{u^-}
\frac{\cB^{(+)+}}{\cB^{(-)+}}\frac{\tilde{\cR}_{1}^{-}\tilde{\cB}_{3}^{-}}{\tilde{\cR}_{1}^{+}\tilde{\cB}_{3}^{+}}\mathcal{D}\right)^{-1} \,.
\label{Wsl2}
\end{multline}
The eigenvalues of the higher double-row transfer matrices in the $sl(2)$
grading ($\tilde{d}_{a,1}$) are obtained by taking into account also the normalization of
(\ref{duallambda}) with $\Lambda^{sl(2)}=\tilde{d}_{1,1}$: 
\begin{eqnarray}
\mathcal{W}_{sl(2)}^{-1} &=& 
\sum_{a}(-1)^{a}\mathcal{D}^{a}\hat{d}_{a,1}\mathcal{D}^{a}\, ,\qquad 
\tilde{d}_{a,1}=h^{[a-1]}h^{[a-3]}\dots h^{[3-a]}h^{[1-a]}\hat{d}_{a,1},
\nonumber \\  
h &=& \rho_1
\left(\frac{x^{+}}{x^{-}}\right)^{m_{1}-2m_{2}}\frac{\cR^{(+)-}}{\cR^{(+)+}} \,.
\end{eqnarray}

Actually it is not difficult to see that the generating functions $\mathcal{W}_{sl(2)}$ and 
$\mathcal{W}_{su(2)}$ (\ref{eq:WWW}) are equivalent. In order to correctly compare 
them, we have to normalize both to generate $\tilde{d}_{a,1}$:
\begin{align*}
\tilde{\mathcal{W}}_{sl(2)}^{-1} &=
\sum_{a}(-1)^{a}\mathcal{D}^{a}\tilde{d}_{a,1}\mathcal{D}^{a}=
(1-\mathcal{D}\tilde{W}_1\mathcal{D})^{-1}  (1-\mathcal{D}\tilde{W}_2\mathcal{D})
(1-\mathcal{D}\tilde{W}_3\mathcal{D})(1-\mathcal{D}\tilde{W}_4\mathcal{D})^{-1} ,
\\  
\tilde{\mathcal{W}}_{su(2)}^{-1} &=
\sum_{a}(-1)^{a}\mathcal{D}^{a}\tilde{d}_{a,1}\mathcal{D}^{a}=
(1-\mathcal{D}{W}_1\mathcal{D})  (1-\mathcal{D}{W}_2\mathcal{D})^{-1}
(1-\mathcal{D}{W}_3\mathcal{D})^{-1}(1-\mathcal{D}{W}_4\mathcal{D}) \,.
\end{align*}
The equality
\begin{equation}
\tilde{\mathcal{W}}_{sl(2)}^{-1} = \tilde{\mathcal{W}}_{su(2)}^{-1}
\end{equation} 
follows from
\begin{equation}
(1-\mathcal{D}\tilde{W}_1\mathcal{D})^{-1}  (1-\mathcal{D}\tilde{W}_2\mathcal{D})=
(1-\mathcal{D}{W}_1\mathcal{D})  (1-\mathcal{D}{W}_2\mathcal{D})^{-1} 
\,,
\end{equation}
and similarly for the other two factors. After inverting the operators this boils down to check that 
$W_1+\tilde{W}_1=W_2+\tilde{W}_2$ and $W_1^{-} \tilde{W}_1^{+}=W_2^{-} \tilde{W}_2^{+}$, which can be easily verified 
from \eqref{id1}, \eqref{id2}.
The identity between two generating functionals shows that any state can be described equivalently by $sl(2)$ and $su(2)$ gradings.
Different forms of the generating functional are related to different 
paths by which the 
B\"acklund transformation can trivialize the $T$-system of $SU(2\vert 2)$. 

Finally we note that in the periodic case the transfer matrices in the $sl(2)$ and $su(2)$ grading are 
related not only by the duality transformation. The transformation which exchanges the labels 
$1\leftrightarrow 3$ and $2\leftrightarrow 4$ and complex conjugates the scattering amplitudes 
is a symmetry of the $SU(2\vert 2)$ scattering matrix. It changes the normalization $S_{11}^{11}=1$
to $S_{33}^{33}=1$ and relates the transfer matrices in $sl(2)$ and $su(2)$ grading as 
\begin{equation}
T^{sl(2)}_{a,s}(p,\{p_i\})=T_{s,a}^{su(2)}(p,\{p_i\})^{\star} \,.
\end{equation}
As the $SU(2\vert2)$ reflection factor transforms under this transformation as 
\begin{equation} 
(e^{-i\frac{p}{2}},-e^{i\frac{p}{2}},1,1)\to (1,1, 
e^{i\frac{p}{2}},-e^{-i\frac{p}{2}}) \,,
\end{equation}
the analogous symmetry relates the double-row
transfer matrices of two different boundary conditions
establishing a kind of duality between them: the symmetric transfer matrices of one 
boundary condition are related to the anti-symmetric transfer matrices of the other.

\section{Computing the sum of residua for the L\"uscher 
correction}\label{sec:trick}

It is not entirely trivial to derive eq.(\ref{egypiper4}) 
as Maple is unable to sum up the residua for $L=2$ and 
$p=\frac{\pi}{4}$ or $p=\frac{3\pi}{4}$. We obtained 
the L\"uscher corrections for these cases in the following way.

After extending the integration domain to the whole real line we use the upper
$q$ plane for the residua. Here, 
in all cases, the integrand has poles on the imaginary axis at $q=ia$ and also 
four others on
two vertical lines at $q=\pm 2u+i(a-1)$ ; and $q=\pm 2u+i(a+1)$. The sum of 
these four residua (at $a$ fixed) can be written as $h(a)-h(a+1)$ for some function $h$, 
with $h(1)=0$. Therefore when we compute the sum over $a$ the residua away from 
the imaginary axis give zero.
The residuum at $q=ia$ is decomposed into partial fractions, those trivially sum up to the $\zeta$-s plus 
the rational part. The rational part is nothing but $h_r(a)-h_r(a+1)$ for some function $h_r$, 
with $h_r(1)=0$. Thus the rational part also vanishes after summing over $a$.

As mentioned above, the rational part is a bit bulky for $L=2$ and $p=\pi /4$ or $p=3\pi /4$. 
In order to decompose them into partial fractions, we notice that they assume the form
\begin{equation}
{\rm Rat} (a) = \sum_{j=1}^J \left( \frac{c_j^{(1)} }{(a - \frac12 - iu)^j }
+ \frac{c_j^{(2)} }{(a - \frac12 + iu)^j }
+ \frac{c_j^{(3)} }{(a + \frac12 - iu)^j }
+ \frac{c_j^{(4)} }{(a + \frac12 + iu)^j } \right),
\label{Rata}
\end{equation}
where $J=2L+7$ for the $1\dot 1$ particle \eqref{egyegy} and $2L+3$ for the $2\dot 2$ particle \eqref{kettoketto}. There should be no constant part to guarantee the convergence of the sum over $a$.
The coefficients $c_j^{(k)}$ can be fixed by using series expansion of the left hand side at each pole. It turns out $c_j^{(1)} = - c_j^{(2)} = - c_j^{(3)} = c_j^{(4)}$ and \eqref{Rata} is rewritten as $h_r(a)-h_r(a+1)$. We also find $h_r(1)=0$ from explicit computation.

\end{document}